# PRIMAD-LID: A Developed Framework for Computational Reproducibility

Meznah Aloqalaa[1], Stian Soiland-Reyes[1] & Carole Goble[1]

0000-0002-1382-3896   0000-0001-9842-9718   0000-0003-1219-2137

## Abstract

Over the past decade, alongside increased focus on computational reproducibility, significant efforts have been made to define reproducibility. However, these definitions provide a textual description rather than a framework. The community has sought conceptual frameworks that identify all factors that must be controlled and described for credible computational reproducibility. The **PRIMAD** model was initially introduced to address inconsistencies in terminology surrounding computational reproducibility by outlining six key factors: **P** (Platforms), **R** (Research objective), **I** (Implementations), **M** (Methods), **A** (Actors), and **D** (Data). Subsequently, various studies across different fields adopted the model and proposed extensions. However, these contributions remain fragmented and require systematic integration and cross-disciplinary validation. To bridge this gap and recognising that PRIMAD provides a broadly applicable framework for reproducibility in computational science, this work undertakes a focused investigation of the PRIMAD model. It combines the model's previous extensions into a unified framework suitable for diverse research contexts. The result is **PRIMAD-LID**, a discipline-diagnostic reproducibility framework that retains the original six PRIMAD dimensions and enhances each with three overarching modifiers: **L**ifespan (temporal qualifier), **I**nterpretation (contextual reasoning), and **D**epth (necessary granularity), establishing a more cohesive and robust foundation for computational reproducibility practices.

---

[1] The University of Manchester, Department of Computer Science, Manchester, M13 9PL, UK

## Introduction

Reproducibility is a fundamental aspect of scientific research, ensuring the validity of findings and enabling other researchers to replicate and expand upon them with confidence. The term 'reproducible research' first appeared in the early 1990s, introduced by geophysicist Jon Claerbout (Claerbout & Karrenbach, 1992). Since then, numerous definitions have been introduced in the literature (Donoho et al., 2009; Peng, 2011; Stodden, 2014; Goodman et al., 2016). A notable consensus report by NASEM, the U.S. National Academies of Sciences, Engineering, and Medicine (National Academies of Sciences, 2019) defines computational reproducibility as the ability to "obtain consistent computational results using the same input data, computational steps, methods, code, and conditions of analysis," and replicability as the ability to "obtain consistent results across studies aimed at answering the same scientific question, each of which has obtained its own data."

NASEM's definitions implicitly invoke a principle of fixity in discussions of reproducibility. Essentially, they acknowledge that certain aspects of an experiment should remain fixed, while others may be allowed to change when results are compared. Additionally, these definitions clearly state the outcomes that should be achievable; they remain primarily textual and, on their own, do not specify an operational recipe for reproducibility in practice. Concretely, reproducibility often depends on publishing and preserving a set of artifacts and decisions such as: (1) versioned source code and a tagged release (e.g., Git commit hash and release archive), (2) a container or virtual machine image capturing the execution environment (e.g., Docker/Singularity recipe), (3) exact parameter and configuration files used for each run (e.g., YAML/JSON configs for model training and inference) and (4) data provenance and versioning (dataset identifiers, checksums, and preprocessing steps).

The PRIMAD framework explicitly operationalises the same underlying notion of "fixity" by systematically characterising computational experiments in terms of six core components: Platform, Research Objective, Implementation, Method, Actor, and Data (Freire et al., 2016). In other words, NASEM provides outcome-focused textual definitions of reproducibility and replicability, whereas PRIMAD provides a component-based framework that makes it explicit what exactly must be held fixed or allowed to vary to substantiate those claims in practice.

# The Original PRIMAD Model

The model, first formulated during Dagstuhl Seminar 16321 in 2016[2], helps researchers reason about the terminology of reproducibility they are evaluating. Within PRIMAD, one specifies which components are held constant and which are varied in a given reproduction attempt, thereby making explicit the conditions under which results are reproduced. This structured approach directly addresses the confusion caused by the inconsistent use of "R words" (De Roure, 2014). By mapping each term to a distinct combination of fixed and variable components of the experiment, the PRIMAD model disambiguates their meanings.

The acronym 'PRIMAD' (Table 1 ) represents the prime variables that can change (or be fixed) during the reproduction of a computational study (Freire et al., 2016):

- **P: Platform:** the execution environment and computational context encompassing hardware, software, compilers, storage, and/or different-purpose platforms (such as management and dissemination).
- **R: Research Objectives** or goals of the study.
- **I: Implementation:** the computational code or source code.
- **M: Methods:** the algorithm, pseudocode, and/or the methodology that fulfil the study goals.
- **A: Actor:** the people who contribute to the study.
- **D: Data:** input data and parameter values of the source code.

Computational reproducibility studies are labelled differently in the PRIMAD framework. In Table 1, computational reproducibility labels are clearly linked to their intended gains and the necessary dimensions to achieve those gains. The presented dimensions clarify which elements will change, adjust accordingly, or remain unchanged. (Aloqalaa et al., 2024) adapted and expanded reproducibility labels from the original PRIMAD table (Freire et al., 2016) to offer more precise guidance on reproducibility practices, their purposes, and outcomes as presented in what they call the "*PRIMAD Taxonomy*".  For example, "Reparameterization" or "Recalibration", which require changes to data parameters to evaluate system sensitivity or optimisation. "Relocate", that require changes to execution environment to evaluate experiment portability.

---

[2] https://www.dagstuhl.de/16321

(Freire et al., 2016) introduce non-direct dimensions to the PRIMAD model: **Consistency** and **Transparency**, as they are not explicitly mentioned in the model acronym. The success or failure of a reproducibility study is evaluated by whether the results are consistent with the previous ones, rather than identical. On the other hand, transparency reflects the ability to examine all necessary components to understand the path from the hypothesis to the results.

| Computational Reproducibility Label | P Platform | R Research Objective | I Implementation | M Method | A Actor | D Data Raw data | D Data Parameters | Reproducibility Gain |
|---|---|---|---|---|---|---|---|---|
| Repeat | - | - | - | - | - | - | - | Consistency Across Trials, Determinism |
| Port, Relocate | X | - | - | - | - | - | - | Cross-Platform Compatibility (Portability), Minimal Dependency (Flexibility) |
| Re-use/Re-purpose | - | X | - | - | - | - | - | Cross-Disciplinary Application (Apply code in different sitting), Resource Efficiency |
| Re-code, Reinterpret | (X) | - | X | - | - | - | - | Improved Code Quality, Correctness of Implementation, Expand Adoption, Enhanced Efficiency, flexibility |
| Ratify, Validate | (X) | - | (X) | X | - | (X) | (X) | Hypothesis Correctness, Validation via a Different Approach, Findings Robustness |
| Review, Independent Verify | - | - | - | - | X | - | - | Enhanced Transparency (Sufficiency information), Independent Verification |
| Resample, Generalize | - | - | - | - | - | X | (X) | Applicability across Different Settings |
| Reparameterization, Recalibration, Parameter Sweep, | - | - | - | - | - | - | X | Robustness, Adaptation to other Conditions (Sensitivity), Parameters Optimization |

Table 1. PRIMAD Taxonomy. The PRIMAD model for reproducibility in a computational experiment; each reproducibility label is associated with its gain and the way of achieving it. What dimensions will change X, change accordingly (X), or will not change - (adapted from (Freire et al., 2016; Aloqalaa et al., 2024).

## Previous PRIMAD Model Adoption

To gain a broader perspective on how the PRIMAD model has been adopted and extended across diverse domains, we conducted a chronological review of the relevant literature. Table 2 presents the findings of this review, listing each study by publication year, domain of application, principal purpose for employing PRIMAD, and whether any extensions to the model were introduced. This consolidated view highlights the varied ways in which PRIMAD has supported reproducibility as a framework in practice and offers insights into its ongoing evolution.

As shown in Table 2, the PRIMAD model has been adopted across diverse domains, ranging from information retrieval (IR) and music IR to biomedical workflows and high-performance computing in astrophysics, visualisation, and even human–computer interaction; as a baseline framework for reasoning about reproducibility. Researchers typically employ PRIMAD to (1) assess and measure reproducibility, (2) identify barriers or gaps, (3) develop tools and guidelines, or (4) refine the model itself. For instance, Ferro et al. (2016) and Ferro et al. (2019) used PRIMAD to measure reproducibility in IR experiments, systematically examining how different experimental components affect outcomes when rerun. While Peter Knees et al. (2022) applied PRIMAD to user-centric MIR experiments, they identified missing elements related to user behaviour and highlighted broader reproducibility barriers. Fekete et al. (2020 also explored reproducibility challenges within the visualisation community, using PRIMAD to pinpoint gaps in documentation and evolving software tools.

In the realm of reproducibility tools, Samuel & König-Ries (2022) blended PRIMAD with semantic provenance ontologies, creating a provenance-based approach to automate completeness checks. In other cases, researchers integrated PRIMAD with complementary frameworks—for instance, aligning it with a trust model for data integrity (Schreyer et al., 2025) or the Whole Tale (Brinckman et al., 2018) concept for computational research objects (Chapp et al., 2020). On the other hand, Staudinger et al. (2024) assessed the PRIMAD model by conducting a "stress-test" in machine learning and information retrieval contexts, systematically varying the model dimensions and ultimately proposing policy- and infrastructure-based enhancements without altering the core PRIMAD framework.

In the following section, we concentrate on studies that directly refined PRIMAD or introduced new extensions. These studies lay the foundation for this work's motivation to extend the model further and address limitations uncovered in prior applications.

| Study (Year) | Discipline | Sub-Discipline | Purpose of PRIMAD Usage | Extension Proposed |
|---|---|---|---|---|
| (Ferro et al., 2016) | Computer Science | Information Retrieval | -Assess Reproducibility | - |
| (Gryk & Ludäscher, 2017) | Life Science | Bioinformatics | -Identify Reproducibility Barriers | - |
| (Freire & Chirigati, 2018) | Computer Science | Computational Science (General) | -Extend/Refine PRIMAD Model | Yes, added Coverage and Longevity to PRIMAD. |
| (Ferro et al., 2019) | Computer Science | Information Retrieval | -Assess Reproducibility | - |
| (Chapp et al., 2019) | Astrophysics | High-Performance Computing | -Assess PRIMAD Model | Partial Refinements, distinguished Method vs. Implementation, noted challenges with Actor, stressed dataset differences, and urged domain-specific refinements. |
| (Chapp et al., 2020) | Astrophysics | High-Performance Computing | -Develop Reproducibility Tools/Guidelines | - |
| (Fekete et al., 2020) | Computer Science | Data Visualization | -Identify Reproducibility Barriers | - |
| (Breuer, 2020) | Computer Science | Information Retrieval | -Assess Reproducibility | - |
| (Schaible et al., 2020) | Computer Science | Digital Libraries and Information Retrieval Platforms | -Identify Reproducibility Barriers -Compare Domain | No formal extension, noted the missing "user" element. |
| (Breuer et al., 2022) | Computer Science | Information Retrieval | - Develop Reproducibility Tools/Guidelines | - |
| (Breuer, 2023) | Computer Science | Information Retrieval | -Extend/Refine PRIMAD Model -Develop Reproducibility Tools/Guidelines | Yes, proposed PRIMAD-U, adding "User" as a new component. |
| (P Knees et al., 2022) | Computer Science & Musicology | Music Information Retrieval | -Identify Reproducibility Barriers | No formal extension pointed out a gap regarding user factors. |
| (Samuel & König-Ries, 2022) | Computer Science | Computational Science | -Develop Reproducibility Tools/Guidelines | - |
| (Aloqalaa et al., 2024) | Life Science | Bioinformatics | -Assess Reproducibility -Extend/Refine PRIMAD Model | Yes, found omissions in PRIMAD, recommended expansions including "Dimension Depth" |
| (Kern, 2024) | Computer Science | Information Retrieval and Machine Learning | -Identify Reproducibility Barriers | - |
| (Staudinger et al., 2024) | Computer Science | Information Retrieval and Machine Learning | -Stress-Test PRIMAD | - |
| (Schreyer et al., 2025) | Computer Science | Human Computer Interaction and Data Management | -Develop Reproducibility Tools/Guidelines | - |

Table 2. Chronological Overview of PRIMAD Model in the literature, its applications, purposes, and extensions across disciplines.

# PRIMAD-LID Motivation and Construction

In the reproducibility literature, it is widely recognised that reproducibility is not a binary concept but rather a spectrum (Peng, 2011; Goodman et al., 2016; Tatman et al., 2018; Gundersen, 2021; Feger & Woźniak, 2022; Raghupathi et al., 2022). Despite this acknowledgement, the numerous definitions of reproducibility in circulation frequently lack consistency, creating obstacles to achieving reproducible experimental results (FASEB, 2016). Several scholars have tackled these semantic challenges from different angles. For instance, Plesser (2017) offered a historical review of 18 definitions of reproducibility published between 1992 and 2017, while Lorena (2018) categorised 43 definitions into distinctions between reproducibility and replicability. Lastly, Gundersen (2021) advanced understanding of computational reproducibility through a systematic survey of 35 scholarly sources (1992–2020), as illustrated in Figure 1.

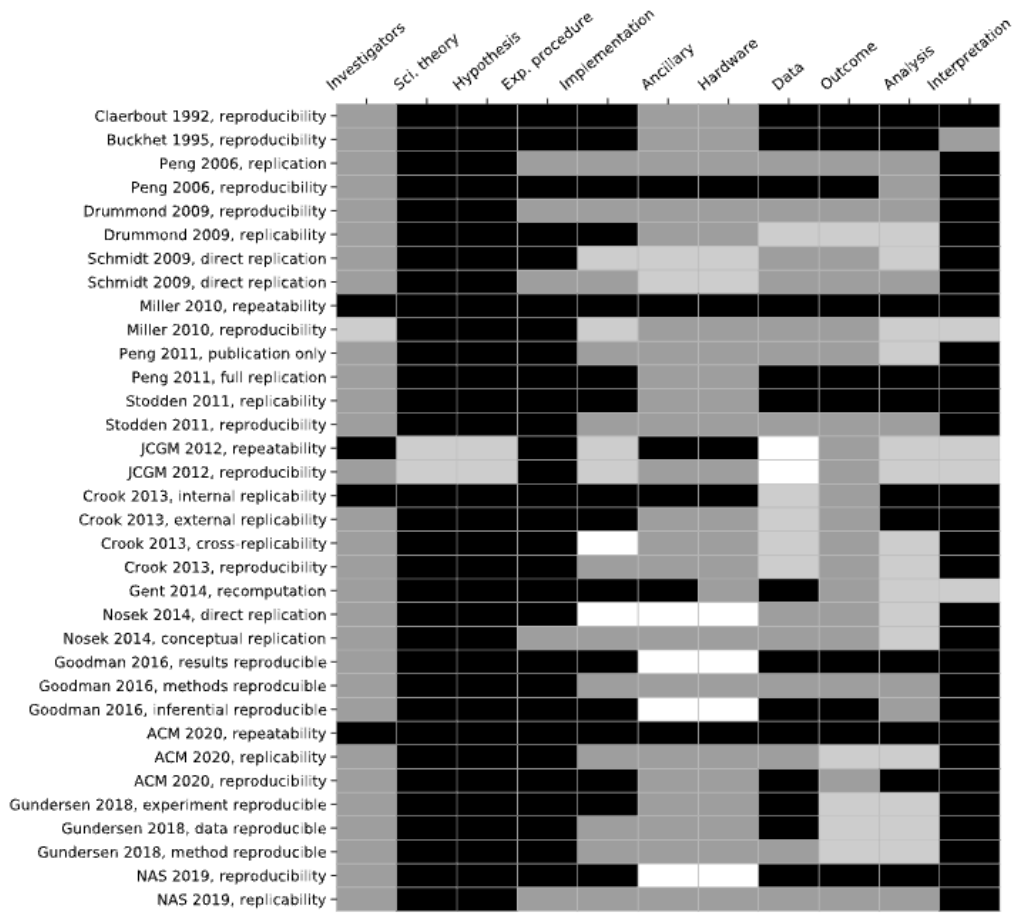

Figure 1. Gundersen Reproducibility Definition Terms Survey: the colour of each cell indicates what has to be kept the same (black), similar (white), different (dark grey) or not specified (light grey) between the original and the reproducibility experiments. Adopted as it is from (Gundersen, 2021)

From that survey, Gundersen identified 11 factors influencing reproducibility: **Investigators, Science Theory**, **Hypothesis**, **Experimental Procedure**, **Implementation**, **Ancillary Software**, **Hardware**, **Data**, **Outcome**, **Analysis**, and **Interpretation**. Each factor was characterised by whether it remained the same, changed, or was unspecified in the definitions under review. Gundersen subsequently defined reproducibility as "the ability of independent investigators to draw the same conclusions from an experiment by following the documentation shared by the original investigators." Notably, "conclusions" are determined by three terms—Outcome, Analysis, and Interpretation—where Outcomes refer to the experiment's output, and Interpretation derives from the outcome analysis (Gundersen, 2021). Based on these distinctions, he proposed three degrees of reproducibility: outcome reproducibility, analysis reproducibility and interpretation reproducibility. Gundersen's specification is more cohesive and accurate than many earlier definitions in determining the degree of the reproduction process, thereby differentiating the results themselves, analysing them, and interpreting them. However, it remains primarily textual and lacks a structured framework.

In contrast, the PRIMAD model offers a more comprehensive, concept-driven perspective, illustrating how various reproducibility terms map onto the key components of a computational study. As noted earlier, researchers have applied the model dimensions in various domains to address specific computational reproducibility objectives (see Table 1). Moreover, some studies have refined or added new dimensions for specialised applications or for more general use, further strengthening the model.

| | PRIMAD Model Dimensions (Freire et al.,2016) | Implicit* | | | Transparency | | | | | Consistency | |
|---|---|---|---|---|---|---|---|---|---|---|---|
| **Original Model** | | Explicit* | Input | Parameter | Implementation | Method | Platform | Actor | Research Objective | | |
| | | | Data | | | | | | | | |
| **Reproducibility Definition Survey Terms** | (Gundersen 2021) | | Data | Outcome | Implementation | Science Theory | Hardware | Investigators | Hypothesis | Analysis | Interpretation |
| | | | | | | Experimental Procedure | Ancillary Software | | | | |
| **Extensions and Refinements to the PRIMAD Model in the literature** | (Chapp et al., 2019) | | Dataset Differences | | | | | Actor Role | | | |
| | (Schaible et al., 2020) (Knees et al., 2022) (Breuer 2023) | | | | | | | End-User or Participant | | | |
| | (Freire and Chirigati 2018) | | Longevity | | | | | | | | |
| | | | Coverage | | | | | | | | |
| | (Aloqalaa et al.,2024) | | Dimension Depth | | | | | | | | |

Table 3. Overview of PRIMAD Model Extensions, mapped to Gundersen's eleven reproducibility factors. This integrated view serves as a roadmap for the proposed extension in this work.

Table 3 summarises these additions and maps them to the original PRIMAD model and Gundersen's eleven reproducibility factors, thus offering a comprehensive view of the model's evolution for our work. Moreover, aligning the model with Gundersen's terminology survey reinforces (Freire et al., 2016)'s recommendation for a deeper examination of PRIMAD's dimensions—and how they vary—to account for the diverse reproducibility definitions adopted across research communities.

In addition to its six explicit dimensions, Freire et al. (2016) introduced **Transparency** and **Consistency** as implicit dimensions. While PRIMAD's core components describe what to reproduce, **Transparency** concerns *how* openly and clearly those components are documented, ensuring that anyone can "look into all necessary components to understand the path from hypothesis to results." This focus arose from realising that even if all PRIMAD components are available, a lack of clear documentation for each one can impede reproducibility. **Consistency,** on the other hand, augments the PRIMAD model by shifting the evaluative criterion from *identical* to *consistent* outcomes. Gundersen (2021) distinguished reproducibility along three degrees—Outcome, Analysis, and Interpretation—which together form the conclusion of a computational experiment. Within PRIMAD, **Analysis** and **Interpretation** align with Consistency, while **Outcome** falls under the Data dimension as part of the experiment "data output", as shown in Table 3.

Later, Freire & Chirigati (2018) proposed **Longevity** and **Coverage** as additional dimensions and emphasised the importance of **Transparency**. As indicated in Table 3, these implicit dimensions are embedded within PRIMAD's six explicit dimensions. **Longevity** emerged from recognising that experiments must remain reproducible well beyond their initial creation, tackling challenges such as software decay, deprecated dependencies, or lost data. Meanwhile, **Coverage** builds on earlier ideas by Freire et al., (2012) addressing how much of an original experiment can feasibly be reproduced. Practical constraints such as proprietary or sensitive data, resource limitations, and the passage of time can result in partial reproducibility. Even with complete transparency, certain experimental steps may be infeasible to replicate, so measuring coverage acknowledges both full and partial reproducibility.

Chapp et al. (2019) investigated the use of PRIMAD in large-scale scientific workflows, with a particular emphasis on LIGO's gravitational-wave data analysis. They introduced partial refinements to the model by **distinguishing between Method and Implementation**, noting that even minor alterations to code or parameters can affect reproducibility without changing the overarching methodology. They also addressed the challenges related to documenting the **Actor dimension**, especially when multiple teams or evolving personnel are involved. Additionally, they stressed **dataset differences**, noting that variations in data type or quality can significantly affect reproducibility. Finally, they urged **domain-specific refinements** to better tailor PRIMAD to the unique needs of high-performance computing and astrophysics research.

In interactive information systems and user studies, **end-users or participants** play an equally critical role alongside researchers. Schaible et al. (2020) were among the first to emphasise the importance of real-user involvement in reproducible IR systems, noting that PRIMAD's Actor dimension should include not only researchers but also participants and end-users who provide feedback. Similarly, P Knees et al. (2022) highlighted the need to reproduce user-centric experiments in music retrieval, emphasising that reproducibility must extend beyond controlled lab settings to account for the inherently variable nature of human behaviour. Breuer (2023) advanced this perspective by introducing **"PRIMAD-U,"** which adds a dedicated "User" dimension to the original six factors. This additional component addresses the varying characteristics and behaviours of end-users, illustrating, for instance, how reproducibility might change if an IR experiment is repeated with a different user group or a simulated user cohort. However, the original PRIMAD model addresses that the Actor dimension includes any person involved in the experiment.

The recent refinement proposed by Aloqalaa et al. (2024) is the concept of **Dimension Depth**. They examined reproducibility in bioinformatics pipelines using PRIMAD with the BioCompute Object (BCO) standard (Simonyan et al., 2017; Alterovitz et al., 2018), an IEEE 2791-2020 standard for bioinformatics analyses generated by high-throughput sequencing (HTS). This work demonstrated that while PRIMAD offers a robust, high-level structure for workflows, it requires more granular specifications within each dimension. In analysing a genomic sequencing workflow, the researcher encountered gaps and ambiguities when mapping PRIMAD onto BioCompute's extensive metadata, revealing the need for deeper detail in each PRIMAD component. Dimension depth extends every PRIMAD dimension through ten cross-cutting attributes: *Time, Additional Resources, Data Categorisation, Functionality, Version, Human Roles, Licence, Fault Tolerance, Review,* and *Metadata Schema*. Thus, PRIMAD is transformed from a set of factors into a richer list. Each dimension effectively becomes a header with detailed fields, enhancing the model's exhaustiveness and making it more discipline-specific.

Consequently, as illustrated in Table 3, effectively integrating implicit dimensions, such as **Consistency** and **Transparency** (Freire et al., 2016), **Longevity** and **Coverage** (Freire & Chirigati, 2018), and **Dimension Depth** (Aloqalaa et al., 2024), along with additional refinements like the **User/Participant** component (Schaible et al., 2020; P Knees et al., 2022; Breuer, 2023) and recommendations from Chapp et al. (2019), requires a systematic organisation of these suggestions within the model's primary dimensions. This process also involves integrating (Gundersen, 2021) **Reproducibility Survey Terms**, thereby ensuring a cohesive framework that encompasses a wide range of reproducibility factors.

## PRIMAD-LID Framework

We propose "PRIMAD-LID" as an enhancement of the original PRIMAD model. The new acronym maintains the six core dimensions of PRIMAD: **P**latform, **R**esearch objective, **I**mplementation, **M**ethod, **A**ctor, and **D**ata, while enriching each dimension with three cross-cutting modifiers: "**L**: Lifespan," "**I**: Interpretation," and "**D**: Depth". As a result, the additional "**LID**" modifiers are aligned perpendicularly to the original PRIMAD dimensions. This alignment provides a more comprehensive and concise framework for examining reproducibility, as will be demonstrated in the subsequent discussion.

### L: Lifespan Dimension

Ensuring software longevity is a critical factor in sustaining long-term computational reproducibility. In addition to the longevity concept proposed by Freire & Chirigati (2018) as an extension of the PRIMAD model, (National Academies of Sciences, 2019) highlighted that obsolescence in data and code storage can pose serious barriers to reproducibility. As operating systems evolve, libraries and dependencies shift, and repositories may be withdrawn or relocated, maintaining software consistency over extended periods becomes a difficult task (Kuttel, 2021). This challenge is particularly acute in research workflows that rely on interconnected tools and data-processing pipelines.

In this extension, we define the **Lifespan dimension (L)** as a *temporal qualifier* for experiment elements, particularly the Method, Platform, Implementation, and Data within PRIMAD. Various temporal indicators, such as issue dates, update dates, version history, and embargo periods, reveal how long each artifact remains valid and functional. By tracking these details, researchers can determine whether, for example, a script or dataset will still be usable several years down the line (Grayson, Milewicz, et al., 2023).

Lifespan dimension can be viewed in terms of *chronological* and *functional* age. Chronological age measures the time elapsed since an artifact's creation, whereas functional age assesses its ongoing usefulness and compatibility. A high chronological age does not necessarily imply obsolescence. For example, the astronomy community continues to use the Flexible Image Transport System (FITS) format, introduced in 1981 (Borgman & Wofford, 2021). Similarly, GitHub has served as a widely adopted code repository for over 15 years, while other services, such as Google Code, have been discontinued despite having similar or even shorter lifespans. On the other hand, functional age depends on factors such as service maturity, community adoption, and continual updates for technical compatibility. Researchers must, therefore, choose platforms and formats with an eye toward sustainability. In their nine best practices for research software registries and repositories, Garijo et al. (2022) recommend implementing *retention* and *end-of-life* policies to support long-term usability.

Real-world studies further illustrate how careful attention to Lifespan enhances reproducibility. (Grayson, Marinov, et al., 2023), for example, examined workflow management platforms (Nextflow (P. A. Ewels et al., 2020)) and Snakemake (Köster & Rahmann, 2012)) with a maximum of five years' worth of revisions. They found that proactively curated updates tripled workflow "wholeness" over time. In the field of natural language processing, repository migrations, broken links, and evolving data standards can reduce the accessibility of datasets and software over time (Mieskes, 2017). There is a continual need to mitigate the effects of the passage of time to facilitate experiment integration and long-term reproducibility. Researchers have highlighted the importance of continuous environment management to address this issue: (Ziemann et al., 2023) refer to the ongoing control of a computational experiment's execution environment as *"compute environment control,"* while (P. Ewels et al., 2019) describe a similar practice under the term *continuous integration* of the computational environment. Both concepts emphasise proactive maintenance of the platform and software components to ensure that changes over time do not compromise reproducibility.

Since the Lifespan of an experiment component cannot be reduced to a single numerical value, we recommend reporting a series of dates: issuing, modification, last access, and review, in addition to version history logs, which reflect its ongoing viability. For instance, the BCO metadata logs an *Obsolescence* date for the entire pipeline, as well as *Embargo, Creation, Modification, Access,* and *Review* dates for each resource (Simonyan et al., 2017; Alterovitz et al., 2018). However, simply logging artifact dates or version numbers is not enough to ensure experimental longevity. Other critical practices, such as establishing open-source environments, systematically documenting data, code, and results in standardised formats, and linking narrative explanations to the analysis (**Interpretation Dimension**), are equally essential. Akhlaghi et al. (2021) proposed eight criteria for computational experiment longevity: *completeness, modularity, minimal complexity, scalability, recorded history, narrative linked to analysis,* and *free/open-source software (FOSS)*. These criteria highlight that reproducibility is not solely about reconstructing the experiment as performed originally; instead, it is about identifying which dimension attributes remain stable and which must be updated or replaced for the experiment's new context or purpose. Within the PRIMAD-LID framework, the Lifespan dimension captures this idea by recognising that each component's longevity can affect reproducibility both individually and in aggregate.

## I: Interpretation Dimension

Scholarship increasingly has recognised the need for an additional dimension of reproducibility: the interpretation of results. The interpretive element captures whether independent researchers arrive at the same conclusions and explanations based on a study's findings (Peng, 2011; Goodman et al., 2016; National Academies of Sciences, 2019; Tatman et al., 2018; Gundersen, 2021). The PRIMAD model was designed to categorise the components of an experiment that can vary during reproduction. However, experts extending PRIMAD stress that reproducibility is *"never a goal in itself"* (Freire et al., 2016); instead, it aims to produce knowledge or evidence that supports scientific claims. For example, changing the method while keeping the same research question allows one to *"validate the correctness of a hypothesis using a different methodological approach,"* providing a method-independent check on the conclusion, referred to as "Ratify" or "Validate" in the *PRIMAD Taxonomy* (Table 1). Essentially, this is an exercise in reproducing the interpretation: if a different method produces consistent results, it supports confidence in the original conclusion.

Additionally, the original PRIMAD framework treated **Consistency** as an implicit dimension, advising that the success of a reproducibility study should be judged by how outcomes will align with the original findings, rather than requiring bit-for-bit identity. Building on this foundation, introducing an **Interpretation dimension (I)** into PRIMAD-LID is a logical step. It encompasses the Consistency and incorporates (Gundersen, 2021) terminology for analysis and interpretation in drawing conclusions, as shown in Table 3. Importantly, Interpretation also accommodates domain-specific contextualisation and justification in reproducibility. It ensures that factors such as the explanation of outcomes, scientific reasoning, and hypothesis validation are explicitly considered. By including an **Interpretation dimension (I)** as a modifier across PRIMAD dimensions, prompting researchers to document and share not only data and code but also the reasoning, decisions, and choices underlying their experiment elements. Moreover, it encourages those replicating the work to report whether they used the same procedures, reached the same conclusions or identified differences in interpretation.

## D: Depth Dimension

The level of detail, or "depth", required to document the experiment components often varies by domain, as different research communities have their own standards for what constitutes reproducible results. The FAIR principles underscore this point, specifying that metadata must include detailed provenance and meet domain-relevant community standards (Wilkinson et al., 2016). Consequently, reproducibility efforts cannot follow a 'one-size-fits-all' approach model (Stewart et al., 2021).

In addition to the original PRIMAD framework (Freire et al., 2016), Freire & Chirigati (2018) emphasise **Transparency** as an implicit dimension that spans all model components. Earlier, Freire et al. (2012) introduced the notion of "depth" to indicate how much information about a computational experiment should be recorded or shared. As mentioned earlier, "depth" was addressed in the context of bioinformatics through the identification of 10 detailed aspects (Aloqalaa et al., 2024). Some domain-specific details are already included in other extended PRIMAD dimensions: *time* and *version* details in Lifespan, *data categorisation* in Data, and *fault tolerance* (variation acceptance) under the Interpretation dimension. Additional aspects, such as *resource details*, *functionality*, *human roles*, *licensing*, and *metadata*, further contribute to capturing the complexity of computational studies more comprehensively.

By adding the **Depth dimension (D)**, researchers can tailor the granularity of their metadata to suit their field's standards, whether by adopting bioinformatics provenance frameworks or machine learning reproducibility checklists. This alignment with community norms enhances the likelihood of successful reproductions. In essence, systematically specifying the depth of information required for each PRIMAD component makes the model more robust and adaptable, and ensures a comprehensive framework for various reproducible computational experiments.

## PRIMAD-LID Coverage

**Coverage** was originally introduced by (Freire et al., 2012) as a characteristic of an experiment regarding reproducibility. Later, an implicit "PRIMAD **coverage**" dimension was proposed to represent the portion of the original experiment's outcomes that is reproducible (Freire & Chirigati, 2018). In essence, coverage corresponds to the reproducible part of the experiment's product or results. This makes coverage a trait of reproducibility itself (i.e., an outcome of applying the model), rather than a descriptive aspect of the experiment's setup. Unlike the core PRIMAD dimensions, which describe how the experiment is conducted or reported (e.g., details about the platform, implementation, methods, data, etc.), coverage indicates reproducibility success, illustrating *how much* of the experiment has been or could be reproduced.

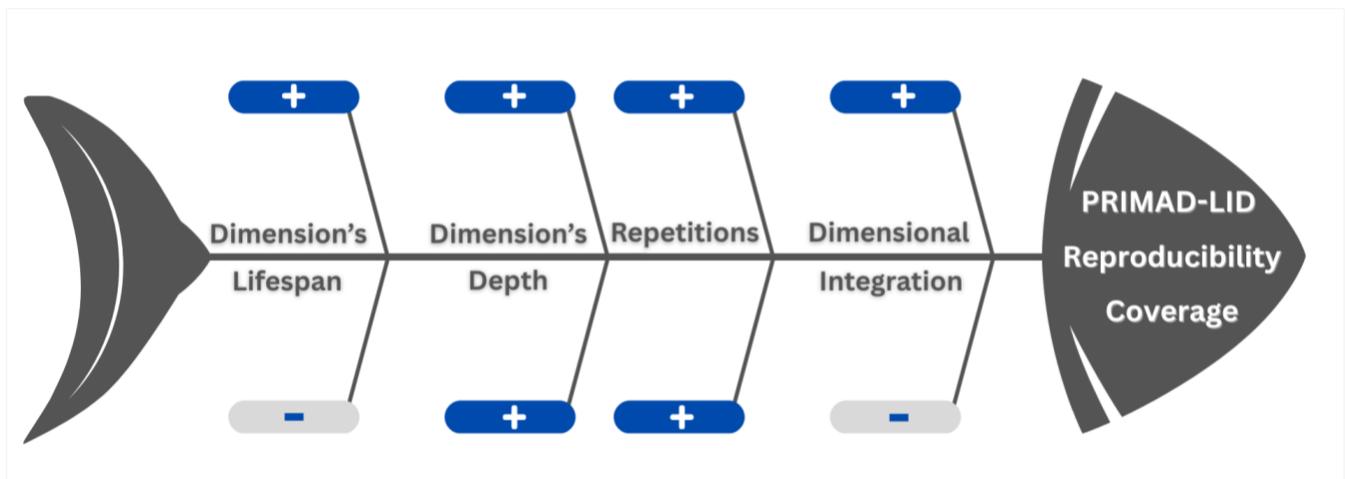

*Figure 2. Influence Diagram for Reproducibility Coverage of the PRIMAD-LID Framework. The diagram illustrates that coverage is positively influenced by the depth of the framework dimensions and the number of previous repetitions for the study. In contrast, the integration between dimensions and their lifespan exerts bidirectional influences (positive or negative) depending on the context.*

The **PRIMAD-LID Coverage** is influenced by three additional factors beyond the original PRIMAD dimensions: **Depth Dimension (D)**, **Lifespan Dimension (L)**, and **Dimensional Integration (T)**. Alongside these, the number of **Repetitions (R)** of an experiment (how often and how widely it is repeated) also critically affects coverage, Figure 2. Each of these factors affects how completely an experiment can be reproduced:

**Depth Dimension (D)**: A higher depth in the model dimension enhances reproducibility coverage by leaving less room for ambiguity. For example, a deeply documented methodology (with thorough descriptions of algorithms, parameters, and protocols) or a richly annotated dataset provides future researchers with clearer guidance, thereby improving the chances that the experiment can be reproduced.

**Lifespan Dimension (L):** Lifespan denotes how long a component of the experiment remains available, usable, or up to date. The impact of lifespan on coverage is nuanced and not uniformly positive or negative; it can vary depending on which dimension is considered (e.g., GitHub). Such longevity can improve reproducibility coverage because the platform (and the materials hosted on it) will likely remain accessible and functional for a long period, allowing others to reproduce the experiment even years later. In contrast, a long lifespan for code components "Implementation" (e.g. scripts or software that remain unchanged for many years) might indicate stagnation, potential obsolescence, or reduced reusability. Empirical evidence suggests caution here: a study on software evolution found that individual lines of code have a median lifespan of about 2.4 years before being modified or removed (Spinellis et al., 2021). This relatively short median lifespan implies that beyond a certain point, code is likely to change, or if it does not change, it may reflect ageing design and dependencies. Thus, a longer lifespan is not inherently good or bad for coverage; its effect depends on context. A stable, well-maintained resource boosts reproducibility, whereas an ageing, unmaintained one might hinder it.

**Dimensional Integration (T):** The PRIMAD model is fundamentally based on the interconnection of an experiment's dimensions, where specifications or changes in one dimension can significantly affect the others (see Table 1). There is often strong interlinking between certain dimensions, specifically among methods, implementations, and platforms (Aloqalaa et al., 2024; Chapp et al., 2019). Integration refers to the degree of coupling or dependency between different experiment dimensions. High integration means the components (data, code, methods, platform, etc.) are tightly interconnected and often packaged or executed in a unified environment. Low integration (or greater distribution) means the components are more independent and can be separated. Tightly integrated experiments (high T) bundle code, data and environment in one self-contained setup, which makes short-term reproduction easy and coverage high, but creates long-term fragility because if the platform or a key dependency break, the whole thing is at risk. Loosely integrated experiments (low T) split these components so that data, code, and environment can evolve, be reused, and be adapted independently over time, which is good for long-term flexibility but makes it harder to fully reconstruct the original experiment later (Akhlaghi et al., 2021). So, integration level shifts the balance: tight boosts immediate reproducibility, loose boosts long-term adaptability, and overall PRIMAD-LID coverage varies accordingly.

**Repetitions(R)** have a clear **positive (+)** relationship with PRIMAD-LID coverage. It refers to how many times the experiment has been repeated and under how many independent circumstances. Each successful repetition, especially by independent researchers or in different contexts, expands the range of conditions under which the results hold, increasing the chance for another successful repetition. Sir Karl Popper famously argued that scientific findings are not credible until they have been tested multiple times: "We do not take even our own observations quite seriously… until we have repeated and tested them. Only by such repetitions can we convince ourselves that we are not dealing with a mere isolated 'coincidence'."(Popper, 2005)

In conclusion, **Coverage** for the PRIMAD-LID model captures the extent to which an original experiment can be reproduced. It is not an absolute property but is influenced by how deeply each component is described, how long each component remains valid or available, how the components are coupled, and how many times the experiment is repeated. A careful balance of depth**,** lifespan, and integration is essential to maximise PRIMAD-LID Coverage. This involves offering rich detail**,** maintaining sustainable resources**,** and selecting an optimal level of integration so that an experiment is both reproducible now and resilient for the future. Additionally, repeating the experiment increases coverage by improving the likelihood that each new replication will succeed, reinforcing the experiment's reproducibility over time.

# Conclusion

Participants in FASEB's roundtable discussions identified the absence of uniform definitions as a major factor impeding the reproducibility of experimental results (FASEB, 2016). Conversely, the UK

Reproducibility Network (UKRN) has emphasised that no single, universal approach can fully address the issue of reproducibility across diverse contexts (Stewart et al., 2021). The PRIMAD model responds to these contrasting recommendations by offering a unified framework that supports multiple reproducibility goals (Freire et al., 2016). Nevertheless, its applications in the literature have demonstrated the need for a more comprehensive extension.

We introduce **PRIMAD-LID**, an expansion of the original PRIMAD model that adds three new dimensions: **Lifespan, Interpretation, and Depth.** These dimensions run perpendicular to the original six (**P**latform, **R**esearch Objective, **I**mplementation, **M**ethod, **D**ata, and **A**ctor), helping clarify and unify the model's definition of reproducibility across various computational science fields. By systematically integrating Lifespan, Interpretation, and Depth, the extended model resolves potential ambiguities arising from refinements to the original framework, ensuring that each newly introduced dimension is clearly defined and practically implementable. It highlights the key components needed to ensure reproducibility and can be adapted to suit different disciplines. Furthermore, PRIMAD-LID provides a modular representation of a computational experiment, crucial for ensuring the experiment's longevity (Akhlaghi et al., 2021). As such, it can serve as a roadmap for computational reproducibility: guiding researchers in their reporting, assisting developers in their workflows, offering publishers clear checklists, and shaping stakeholders' guidelines for research data management. Nevertheless, the PRIMAD-LID framework requires further examination across various computational fields. This is because the *depth* of each dimension can vary significantly both across and within disciplines, depending on the specific methods employed.